# Effect of different Zinc precursors in Structural and Optical properties of ZnO thin films

N.lehraki[a*], A. Attaf[a], M. S Aida[b], N. Attaf[c], M. Othmane[a] and F. Bouaichi[a]

[a]Laboratory of Physics of thin layers and applications, Biskra university, BP 145 Biskra 07000, Algeria

[b]King Abdulaziz University · Department of Physics, Jeddah, Saudi Arabia

[c]University of Constantine 1, Department of Physics, Constantine

**ABSTRACT**

Zinc oxide (ZnO) thin films were deposited by ultrasonic spray pyrolysis technique using different Zinc precursors: Zinc acetate, Zinc nitrate and zinc chloride with different molarities. All deposited films were characterized by various techniques such as X-ray diffraction to determine the films structure, the scanning electron microscopy (SEM) for the morphology of the surfaces and UV–visible spectroscopy to determine the optical proprieties. The obtained results indicated that ZnO nano-crystalline films have a hexagonal structure at type wurtzite for all Zinc precursors. Films deposited with zinc acetate are characterized by a smooth surface, dense network and high transparency, while films deposited with zinc chloride and Zinc Nitrate have a better crystallinity and low optical transmittance.

## 1. INTRODUCTION

Zinc Oxide is an important, inexpensive, versatile n-type semiconducting material with widedirect energy band gap of 3.37 eV and large exciton binding energy of about 60meV at roomtemperature [1]. It is also attractive in the field of semiconductor due to its excellent optical properties, high stability, cheap and abundant element and good electrical properties compared toother materials [2]. It has been used in many electronic applications such as solar cell [3], gas sensors [4], light emitting diodes [5], transistors and microelectronic devices [5] and piezoelectric transducers [5]. Various techniques have been used to produce ZnO thin films, among them the chemical methods which are achieved in a liquid medium based on solutions, prepared from different precursors. The most important techniques using this process are: sol-gel [6,7],electrodeposition [8,9], chemical bath deposition (CBD) [10,11] and spray pyrolysis technique(SPT) [12,13]. Among these methods, ultrasonic spray pyrolysis is one of the effective andattractive techniques for preparing ZnO thin films on inexpensive substrate. The economic viability of SPT derives from theadvantage of inexpensive equipment (non-vacuum method), the ease of large area deposition[14].

In thepresent study ZnO thin films were prepared successfully,on glass substrate,using ultrasonic spray technique with easy processing steps. In this work, we have investigated the effect of three type of Zinc precursors and solution concentration on the structural and optical properties of the ZnO films.

**Corresponding author:** [*]N.Lehraki

**Email: Nadia.lehraki@univ-biskra.dz**

## 2. EXPERIMENTAL PROCEDURE

Three different solutions, with different molarities (0.1, 0.2, 0.3 and 0.4 mol/l) were prepared by mixing zinc salts in absolute volume of methanol $CH_3OH$ as a solvent. The used salts are zinc acetate, zinc nitrate and zinc chloride with 99.9995% of purity. All thin films were deposited on heated glass substrates using ultrasonic spray process. ZnO thin films were deposited with the prepared solutions on well-cleaned glass substrates with Acetone, methanol and distille dwater. In all depositions, the substrates were heated at temperature equals to 400 °C. The distance from the spray nozzle and substrate was fixed at 4.5 cm.

The structure characterization was carried out by X-ray diffraction using a D8 ADVANCE Diffractometer with a $CuK_\alpha$ ($\lambda$=1.5405 °A) radiation for 2θ values in the range of 20-80°. Film surface morphology was investigated using JOEL model JSM6301F a scanning electron microscopy. The optical transmittance spectra were obtained using UV–VIS-NirVarian-Cary spectrophotometer; these measurements were performed using glass as reference in a wavelength range of 200 – 800 nm.

## 3. RESULTS AND DISCUSSION

### 3.1 Surface morphology

Surface morphology and cross section of ZnO samples deposited at growth temperature $T_s$=400°C with different molarities and different Zinc precursors are presented in **Fig. 1**. As can be seen, the film morphology depends strongly on the nature of the used precursors. Film deposited with zinc acetate solution has a continuous and dense structure with a smooth surface morphology as reveals the **Fig. 1(a)**. The film deposited with zinc chloride solution is formed with a separated hexagonal column perpendicular to the substrate surface **Fig.1(b)**.

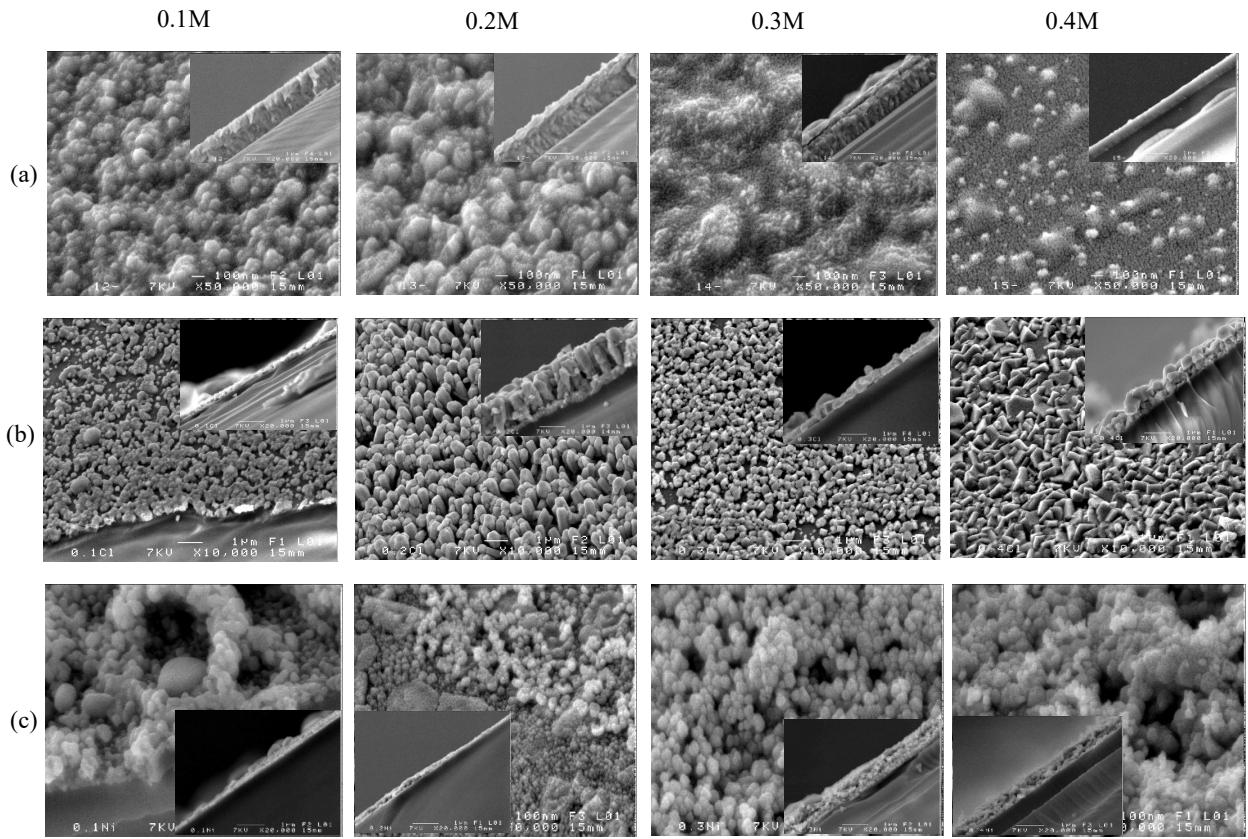

**Fig. 1**. SEM images of ZnO thin films deposited at substrate temperature equal to 400°C with different solutions: (a) Zinc acetate (b) Zincchloride and (c) Zincnitrate

Film prepared with zinc nitrate as starting solution has porous and rough surface morphology **fig. 1(c)**. The dissociation energy also plays a more important role in the formation of thin films; the following table gives the values of the dissociation energy of each source of Zinc.Indeed, the source that has the lowest dissociation energy than the fastest growth rate

| Table1. The dissociation energy of different sources of zinc[15] | | | |
|---|---|---|---|
| **Zinc salts** | **Zinc Acetate** | **Zinc Nitrate** | **Zinc Chloride** |
| **ΔH (enthalpy of surface formation) (Kcal/mol)** | 0.1 | 10 | 30 |

From cross section, it can be seen that for concentrations of 0.1 mol / l, the growth rate to produce ZnO thin film from zinc acetate is high; approximately 130 nm / min, then the ZnO film produced by zinc nitrate20 nm / min and low for the film prepared by zinc chloride.

### 3.2 Structural properties

**Fig. 2** demonstrates XRD spectra of ZnO thin filmsdeposited with the three studied salts and at substrate temperature 400°C taken as example, with different molarities. The comparison of the DRX spectra with the JCPDS data card (36–1451) [16] makes it possible to identify the films structure. All diffractograms exhibit two peaks located at 34.5 ° and 34.8 ° (**Table2**.), These peaks are assigned to the ZnO hexagonal Wurtzite structure with a privileged orientation of the growth along the crystallographic axis c (002) perpendicular to the substrate. It is well argued that ZnO, regardless the deposition technique, tends to grow in the direction (002) due to its low surface energy [17,18]. The other peaks correspond to the other planes (100), (101), (110) and (103) of the hexagonal structure of ZnO.

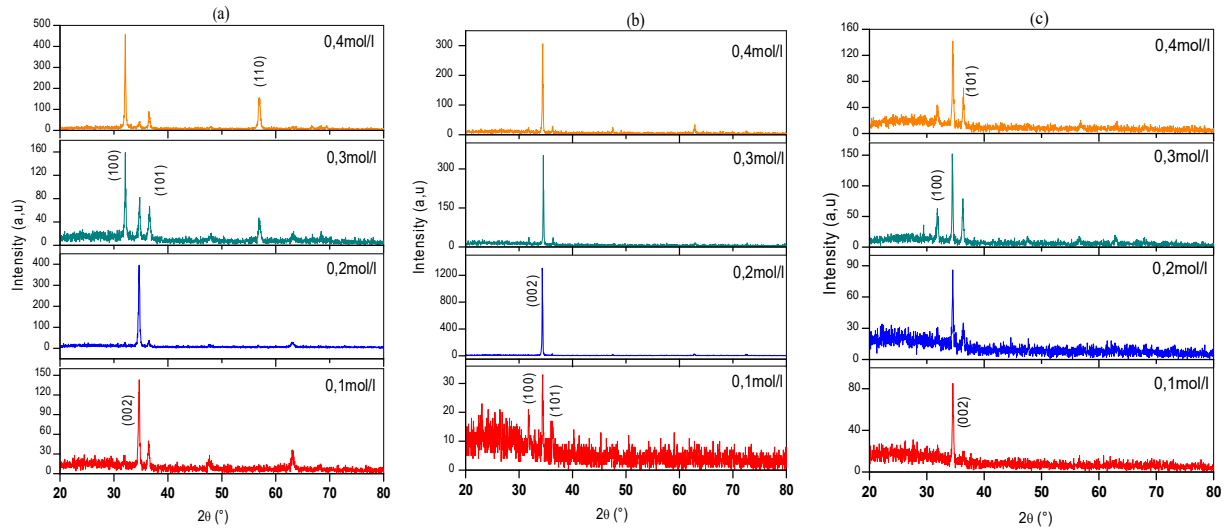

**Fig.2.** XRD patterns of ZnO films prepared with different molar concentration of different Zinc salts: (a) Zinc acetate (b) Zinc chloride and (c) Zinc nitrate

There is an exception for ZnO films prepared from Zinc acetate solution. It has been seen that the preferred orientation of growth is along axis (100) for the molarity greater than 0.2 mol / l. Lee and al [19] have shown that the variation of the orientation could be explained by a reduction in filmthickness, since the probability of rearrangement of the atoms to produce (100) is low at high thickness.In the case of zinc chloride precursor, for the molarity equal to 0.2M, ZnO thin film is well crystallized along the (002) orientation. This is obvious that the crystalline structure seems to be a hexagonal form (**Fig.1(b)**). It is also noticed for the spectra shown in figure2.c,by increasing the molar concentration of precursors, the intensity of preferential orientation (002) increases indicating the films crystallinity improvement [20].

From the predominant peaks in the XRD patterns of ZnO thin films, the structural paramaters are calculated and are summarizedin table2.

| Zinc precursors | Molarity (mol/l) | 2θ(°) | a (nm) | c (nm) | c/a | D(nm) | δx10⁻³ (nm⁻²) | $\varepsilon_z$ | Stress (GPa) |
|---|---|---|---|---|---|---|---|---|---|
| Zinc Acetate | 0.1 | 34,69 | 0.3227 | 0.5174 | 1,603 | 45 | 0,498 | -0.0063 | 1,458 |
|  | 0.2 | 34,66 | 0.3227 | 0.5179 | 1,601 | 78 | 0,163 | -0.0054 | 1,264 |
|  | 0.3 | 34,78 | 0.3224 | 0.5161 | 1,600 | 39 | 0,650 | -0.0087 | 2,039 |
|  | 0.4 | 34,81 | 0.3224 | 0.5157 | 1,599 | 26 | 1,462 | -0.0095 | 2,232 |
| Zinc Chloride | 0.1 | 34,42 | 0.3259 | 0.5214 | 1,60 | 41 | 0,203 | 0.0013 | -0,302 |
|  | 0.2 | 34,33 | -- | 0.5227 | -- | 63 | 0,580 | 0.0038 | -0,895 |
|  | 0.3 | 34,54 | 0.3250 | 0.5196 | 1,598 | 69 | 0,249 | -0.0021 | 0,483 |
|  | 0.4 | 34,42 | -- | 0.5214 | -- | 53 | 0,206 | 0.0013 | -0,302 |
| Zinc Nitrate | 0.1 | 34,54 | 0.3257 | 0.5196 | 1,595 | 42 | 0,551 | -0.0021 | 0,484 |
|  | 0.2 | 34,51 | 0.3245 | 0.5200 | 1,602 | 64 | 0,241 | -0.0012 | 0,288 |
|  | 0.3 | 34,42 | 0.3251 | 0.5214 | 1,601 | 54 | 0,348 | 0.0013 | -0,302 |
|  | 0.4 | 34,54 | 0.3257 | 0.5196 | 1,595 | 41 | 0,603 | -0.0021 | 0,484 |

**Table2.** Structural parameters obtained from the analysis of the predominant peaks in the XRD patterns of ZnO thin films deposited with different Zinc precursors

The value of lattice parameter are calculated using (1) [21]

$$\frac{1}{d_{hkl}} = \left[\frac{4}{3}\left(\frac{h^2+hk+k^2}{a^2}\right) + \frac{l^2}{c^2}\right]^{\frac{1}{2}} \quad (1)$$

The crystallite size (D) of the ZnO films was calculated using theclassical Scherrer formula given by [10]:

$$D = \frac{k\lambda}{\beta \cos\theta} \quad (2)$$

where, the constant k is the shape factor (usually equal to 0.94), λ is the wavelength of X–ray, θ is the Bragg's angle and β is the full width of the half maxima (FWHM). It is well knownthat the crystallite size measured by this method is usually lessthan the actual value.

Strain of the ZnO films along the axis is calculated using "(3)"

$$\varepsilon_z = \frac{c-c_0}{c_0} \quad (3)$$

Where c is the lattice parameter of the ZnO films and $c_0$ is the unstrained lattice parameter of ZnO equalsto 0.5206 nm. The lattice stressin the ZnO thin films is calculated from the following relation[22]:

$$\sigma = \left[\frac{2C_{13}^2 - (C_{11}+C_{12})C_{33}}{C_{13}}\right]\varepsilon_z \quad (4)$$

Here, $C_{ij}$ are the elastic stiffness constants for ZnO. Using the numerical values of $C_{13}$ = 104,2 GPa, $C_{33}$ = 213,8 GPa, $C_{11}$ = 208,8 GPa et $C_{12}$ = 119,7 GPa [23,24] the proportionality constant is calculated as – 233 GPa.

The dislocation density (δ), the dislocation lines per unit area is evaluated using the Williamson and Smallman's formula [25] as executed in (5)

$$\delta = \frac{1}{D^2} \quad (5)$$

It can be concluded from **Table 2**, that the lattice parameters (a) and (c) are related to the Zinc precursor type, for Zinc acetate and Zinc nitrate the values of(c) are less than the value of "$c_0$". The standard value of lattice constants of ZnO crystal are a = 0.32495 nm, and c = 0.52069 nm.These values indicate that the films have undergone a compressive stress along the axis c, as the molarity increases. This compression constrainte usually occurs if there are defaults and distortions in the crystalline network [26]. On the other hand, for Zinc chloride, the deposited layers have undergone tensile stress along the axis perpendicular to the substrate. Stress is maximum for ZnO thin films deposited by Zinc acetate.

In **Table 2**, it is clearly seen that D value lies between 26 and 78 nm which is consistent with the results found for ZnO by other researches [27-30]. In the case of Zinc acetate and Zinc nitrate, D decreases with the increase of molar concentration and increases for Zinc chloride. ε and δ value indicate that the crystallites formed for molar concentration equal to 0.2 mol/l are less strained and with low dislocations are formed in films prepared using Zinc acetate and Zinc chloride precursors, films preparation with Zinc chloride are less strained when using 0.3mol/l concentration.

### 3.3 Optical properties
#### 3.3.1 Transmission

The Optical transmission curves of ZnO thin films recorded as a function of wavelength in the range of 200 nm to 800 nm are shown in **Fig.3**.

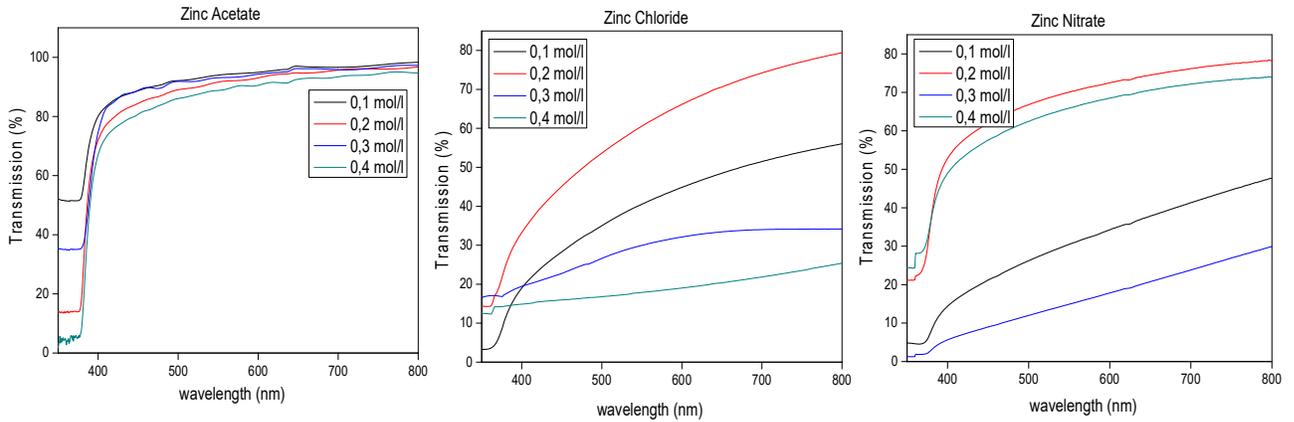

**Fig.3.** Transmission spectra with different Zinc precursors and molar concentration

In the case of Zinc acetate, it is noted that average transmission in the visible range [380-780] between 80 and 95%, on the other hand the transmission of the films obtained from zinc chloride and zinc nitrate are a little less transparent, the films show an average transmission varying between 20-64% and 18-81% respectively. We also saw the absence of interference fringes is due to the roughness of the free surface of our thin films prepared by zinc chloride and zinc nitrate, this roughness causes the scattering of light instead of reflection on the interface.

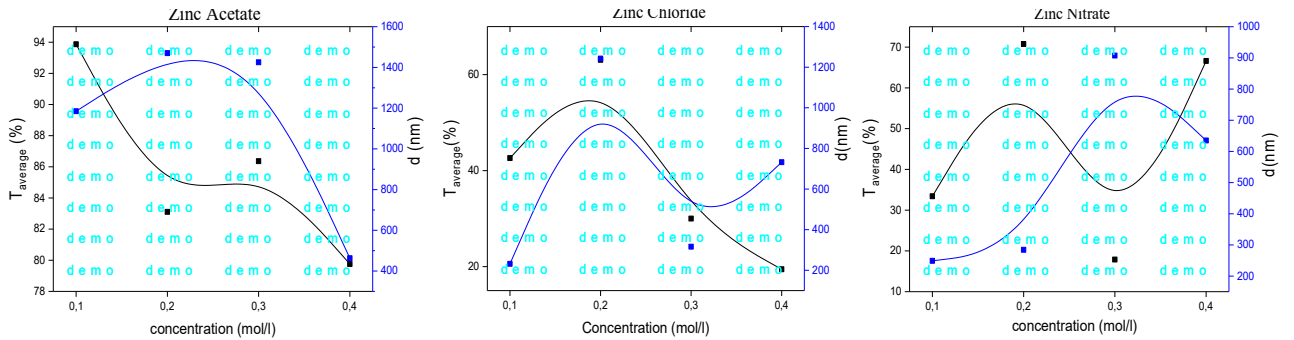

**Fig.4.** Evolution of average Transmission and thickness with different molar concentration

**Fig.4.** show the evolution of average transmission and thickness of our films as function of molar concentration, We resume:

From **Fig.4(a)**, in the case of thin films prepared with zinc acetate, the variation in transmittance can be explained in terms of film thickness variation, knowing that thicker films are less transparent. The reduction of the transmittance to a molar concentration of 0.4 mol / l, although the sample thickness is small, can be attributed to the lower film crystallinity that may causes photons scattering in visible region of solar spectrum [20,28].

**Fig.4(b)** shows the decrease in average transmittance with increasing molarity. As can be seen a high transparency when molar concentration is equal to 0.2 mol/l. It is noticeable that (002) and (110) peaks in the XRD patterns of the ZnO thin film (**Fig.3**.), influences the transmittance value. The transmittance increases with dominance of (002) peak and disappearance of (110) peak. Sharmin [31]indicated that the preferred growth orientation along the c axis of film of ZnO thin films causes enhancement in the value of transmittance and our thin film shows good structure hexagonal with grain size about 60 nm.

**Fig.4(c)** shows a good agreement between the average transmittance and the thickness of films deposited by Zinc nitrate.

The optical band gap of ZnO thin films was calculated using method of Tauc [32]:

$$(\alpha h\nu)^2 = A(h\nu - E_g) \qquad (6)$$

With
$$\alpha = \frac{Ln(T)}{d} \qquad (7)$$

where, α is absorption coefficient, A is the constant independent of photon energy (hν), h is the Planck constant, $E_g$ is the energy band gap of the semi- conductor, d is the film thickness and T is transmittance.
The value of optical band gap can be estimated byextrapolation of the linear region to $(\alpha h\nu)^2 = 0$.
**Fig.5.** shows the band gap energy values of ZnO thin film deposited with three Zinc precursors

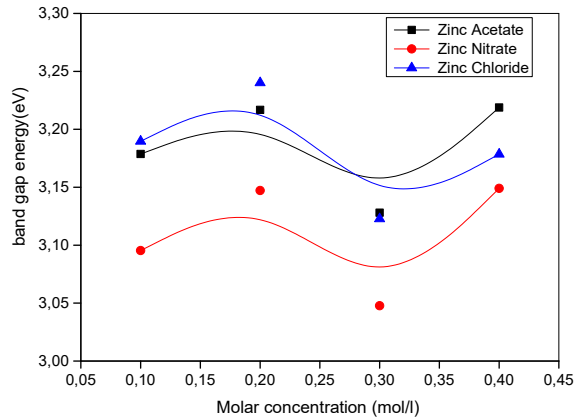

**Fig.5.** Variation of band gap energy as function of molar concentration for different Zinc precursors

The obtained values of the optical band gap of all prepared films are relatively lower than those of single crystal (3.37 eV). The difference in these values is within the experimental error of the used method. Gap energy values of ZnO thin films increases with increasing the molar concentration between 0.1 and 0.2 mol/l and it decreases from 0.2 to 0.3mol/l and rise again from 0.3 to 0.4mol/l.The variation of ZnO films gap energy with the used molar concentration is related to the change in grain size and stress in films as shown in **Table2**.

**4. CONCLUSION**
ZnO thin films were prepared using zinc acetate, Zinc chloride and Zinc nitrate precursors by ultrasonic spray technique on glass substrates at 400°C, with different molar concentration. Regardless the precursors nature, all the prepared ZnO films exhibit the Wurtzite hexagonal structure. The thin film grown using Zinc chloride as precursor with molar concentration 0.2 mol/l have a well preferred orientation of (002), while the other precursors. SEM image reveals that film deposited with zinc chloride exhibits bettercrystallinity. While films deposited with acetate and nitrideprecursors are polycrystalline. Zinc acetate solution yields to film with highertransmittance by comparison to films prepared with

Zinc chloride and Zinc acetate precursors. The calculated band gap of ZnO films using different precursors are lower than the band gap of bulk ZnO crystal and are strongly influenced by the stress.